\documentclass[pdflatex,sn-mathphys-num]{sn-jnl}

\usepackage{graphicx}%
\usepackage{multirow}%
\usepackage{amsmath,amssymb,amsfonts}%
\usepackage{amsthm}%
\usepackage{mathrsfs}%
\usepackage[title]{appendix}%
\usepackage{xcolor}%
\usepackage{textcomp}%
\usepackage{manyfoot}%
\usepackage{booktabs}%
\usepackage{algorithm}%
\usepackage{algorithmicx}%
\usepackage{algpseudocode}%
\usepackage{listings}%
\usepackage{subfigure}
\usepackage{indentfirst}

\usepackage{soul}

\theoremstyle{thmstyleone}%
\theoremstyle{thmstyletwo}%

\theoremstyle{thmstylethree}%

\begin{document}


\title[Article Title]{Discrete scaling in non-integer dimensions }


\author*[1]{\fnm{T.} \sur{Frederico}}\email{tobias@ita.br}
\equalcont{These authors contributed equally to this work.}

\author[1]{\fnm{R. M.} \sur{Francisco}}\email{elfara2311@gmail.com}
\equalcont{These authors contributed equally to this work.}

\author[1]{\fnm{D. S.} \sur{ Rosa}}\email{derick.s.rosa@gmail.com}
\equalcont{These authors contributed equally to this work.}

\author[2]{\fnm{G.} \sur{ Krein}}\email{gastao.krein@unesp.br}
\equalcont{These authors contributed equally to this work.}

\author[2]{\fnm{M. T.} \sur{  Yamashita}}\email{marcelo.yamashita@unesp.br}
\equalcont{These authors contributed equally to this work.}

\affil[1]{\orgdiv{DCTA}, \orgname{Instituto Tecnol\'{o}gico de Aeron\'{a}utica}, \orgaddress{\street{Pr. Mal. Eduardo Gomes}, \city{S\~{a}o Jos\'{e} dos Campos}, \postcode{12228-900}, \state{SP}, \country{Brazil}}}

\affil[2]{\orgdiv{Instituto de F\'isica Te\'orica}, \orgname{Universidade Estadual Paulista}, \orgaddress{\street{Rua Dr. Bento Teobaldo Ferraz, 271-Bloco II}, \city{S\~ao Paulo}, \postcode{01140-070}, \state{SP}, \country{Brazil}}}


\abstract{We explore the effect of a finite two-body energy in the discrete scale symmetry regime of two heavy bosonic impurities immersed in a light bosonic system. By means of the Born-Oppenheimer approximation in non-integer dimensions $(D)$, we  discuss the effective potential of the heavy-particles Schrodinger equation. We study how  including the two-body energy in the effective potential changes the light-particles wave function and the ratio between successive Efimov states. We present the limit cycles associated with correlation between the energy of successive levels for the three and four-body systems. Our study is exemplified by considering a system composed of $N$-bosons, namely two Rubidium atoms interacting with $N-2$ Lithium ones ($^7$Li$_{N-2}-^{87}$Rb$_2$), which represent compounds of current experimental 
interest.}




\maketitle

\section{Introduction}\label{sec1}

Efimov physics is associated with universal properties exhibited by three-body systems in atomic and nuclear contexts~\cite{Braaten:2004rn,
Frederico:2012xh,RevModPhys.89.035006}, as well as in a variety of different
three-body systems~\cite{efimovatoms,efimovdipolar,efimovphotons,
efimovpolarons} in the situation of resonant s-wave short-range interactions. 
In the limit of infinite scattering lengths (unitarity limit) the manifestation of an infinite tower of three-body bound states, when the sub-systems interact by means of resonant short-range two-body forces, is 
known as Efimov effect~\cite{efimov}. This phenomenon appears as 
a universal geometrical scaling law in its spectrum, 
with its first evidences found experimentally in 2006 by Kraemer et 
al.~\cite{kraemer} in the study of trapped ultracold atomic systems close to 
Feshbach resonances (see also~\cite{homoexp1,homoexp2,heteexp1,
heteexp2,heteexp3}). 

A variety of analytical and numerical methods can be used to study 
the Efimov effect, such as the Bethe-Peierls (BP) boundary condition alongside with the Faddeev decomposition~\cite{Naidon:2016dpf}, the 
Skorniakov and Ter-Martirosyan (STM) equations~\cite{skorniakov1957three}, 
and in framework of effective-field theory~\cite{PhysRevA.94.032702}. 
These methods can be used in different contexts and with different purposes.
The BP boundary condition approach allows to solve the Faddeev 
equations analytically to obtain the bound state wave 
function for finite binding energies in the unitarity limit~\cite{rosabp}. 
In the STM framework, the three-body wave function for 
mass imbalanced systems was reviewed by Minlos~\cite{2014RuMaS..69..539M}. 

It is well known that the Efimov effect is present also in
heteronuclear systems. The mass imbalance
modifies the discrete scaling characteristic of Efimov states, 
preserving however the universality of the three-body 
properties. In particular, 
the mass imbalance tends to enhance the attractive Efimov long-range 
$(1/r^2)$ effective interaction in the case of heavy-heavy-light systems, 
which makes the geometrical ratio between successive Efimov energy levels 
small. This is well known and these situations represent the best scenario 
for studying multiple excited Efimov states (see e.g.~\cite{Naidon:2016dpf}). 
In fact, the highly mass-imbalanced system $^6$Li-$^{133}$Cs$_2$ provided the 
most convincing experimental results for the universal Efimov 
scaling~\cite{RevModPhys.89.035006}. 

While exploring the Efimov physics, an important aspect to be considered is 
the deformation of the trap that confines the system. 
To mimic such a deformation, 
three-body systems have
been studied in non-integer dimensions, both for continuous 
and bound states~\cite{mohapatra,cristensen,garridocont,induefimov,rosabp}. 
In particular, it has been theoretically demonstrated that for homonuclear 
systems, the Efimov effect survives only for dimensions in the range $2.3 < D < 3.8$~\cite{nielsen}. It is important to mention that one can relate the effective dimension~$D$ to the
the aspect ratio of the confining trap.
Such a relation was found in Ref.~\cite{garridoprr} by 
comparing the three-body bound state 
energies of a squeezed system by the action of an external harmonic 
potential and the corresponding energies in non-integer dimensions, namely:
\begin{equation}
3(D-2)/(3-D)(D-1)= b_{ho}^2/r_{2D}^2\,,
\label{eq:Dtrap}
\end{equation} 
where  $b_{ho}$ is the oscillator length and $r_{2D}$ is the root mean 
square radius of the three-body system in two dimensions. 
This relation was tested for Gaussian and Morse short-range two-body
potentials, suggesting that it is independent to the details 
of the two-body potential~\cite{garridoprr}. Experimentally, it is possible to compress and expand atomic clouds, creating effectively two-~\cite{BEC2D} and one-dimensional~\cite{BEC1D} setups, but, to the best of our knowledge, changing the trap geometry continuously while controlling the Feshbach resonance is challenging.

Besides the studies performed for three-body systems, various studies 
have been made to investigate the manifestation of  Efimov-like phenomena for 
four or more particles~\cite{kroger,adhikari,naus,yamashita81,
yamashita75,von,yan,NaidonFBS2018}). In 2011, for a system of four bosons 
with contact interaction, it was theoretically discovered the existence 
of a limit cycle for the energies of the tetramers that is independent of 
the Efimov one~\cite{hadizadeh2011} (a recent study can be found in 
Ref.~\cite{Frederico:2023fee}). Following that, the shift in the
four-body recombination peaks due to finite effective range
corrections was studied in Ref.~\cite{hadizadeh2013}. 
More recently, the independence between few-body scales was demonstrated
for $N$-boson systems composed by two heavy atoms interacting 
with ($N-2$) light ones at the unitary limit~\cite{interwoven}, 
showing different limit-cycles, each one associated with a given 
number of bosons present in the system. Following this 
idea, now for a heavy-light system with a resonant s-wave interaction,
Ref.~\cite{prafrancisco} studied the spectrum and 
structure of two heavy bosonic impurities immersed in a light 
boson system in noninteger dimensions $D$. Using
the Born-Oppenheimer approximation, as used earlier for heavy-heavy-light systems with short range 
interactions~\cite{Fonseca:1978kj}, Ref.~\cite{prafrancisco} found
that this system exhibits an Efimov-type geometrical scaling law as a 
function of the mass asymmetry of the system up to a critical dimension 
depending on the mass imbalance and number of light bosons. 

In the present work, we extend the investigation of the mass imbalanced $N$-bosons system of Ref.~\cite{prafrancisco} by studying the case of 
$^7$Li$_{N-2}-^{87}$Rb$_2$, in which, in the adiabatic heavy-heavy potential 
generated by the light particles, we consider a finite heavy-light binding 
energy. We aim to understand how these corrections affect the bound state 
spectrum. To accomplish this, we review the Born-Oppenheimer approach and 
derive the heavy-heavy effective potential at large and short distances. In addition, we compute the spectrum and wave function. 

The work is organized as follows. In section~\ref{sec:BOA}, 
we use the Born-Oppenheimer approximation
to discuss the heavy-heavy effective potential and analyze its asymptotic behavior, besides
studying the limit-cycles 
in the $^7$Li$_{N-2}-^{87}$Rb$_2$ system. 
In addition, we discuss the wave functions for the 
light-particles and for the heavy-heavy pair of particles
in~$D$ dimensions. Section~\ref{sec:conclusion} presents our final considerations. 

\section{BO Approximation: heavy-heavy and light bosons}
\label{sec:BOA}

In what follows, we review the Born-Oppenheimer (BO) approach used to study the mass-imbalanced system composed by two identical heavy bosons with masses $m_A$ and ($N-2$) light bosons with identical masses $m_B$.

For $m_B \ll m_A$, following Ref.~\cite{prafrancisco} in the situation 
where the heavy-light interaction is dominant, we can write separated 
equations for the light and heavy bosons 
\begin{equation}
\left[-\frac{\hbar^{2}}{2\mu_{B,AA}}\nabla^{2}_{r_i}  + \sum_{j=1}^{2}V_{A}\left(\Big\vert\textbf{r}_i+(-1)^{j}\frac{\textbf{R}}{2}\Big\vert\right)\right]\psi_R(\textbf{r}_{i}) = \epsilon(R)\psi_R(\textbf{r}_{i})\,, 
\label{eqlightatom}
\end{equation}
and
\begin{equation}
\left[- \frac{\hbar^{2}}{2\mu_{AA}} \nabla_{R}^{2}+ (N-2)\epsilon(R) \right] \phi(\textbf{R})
 = E_{N} \phi(\textbf{R})\,,\label{eqheavy}
\end{equation}
where $\mu_{AA} = m_{A}/2$ and $\mu_{B,AA} = 2m_A m_B /(2m_A +m_B )$ are the 
reduced masses of the system and $V_A$ denotes the heavy-light potential. The 
energy eigenvalue denoted by $\epsilon(R)$ in  Eq.~\eqref{eqlightatom} works 
as an effective potential in the energy eigenvalue equation for the 
heavy-heavy system, Eq.~\eqref{eqheavy}, where $E_{N}$ is the energy 
which includes the contribution from the light bosons. Note that, 
although not indicated, the kinetic and potential energy operators are 
given in $D$-dimensions.

\subsection{Effective heavy-heavy potential}
\label{sec:EP}

The effective BO heavy-heavy potential for 
a heavy-light contact interaction, 
$V_{AA}(R)=(N-2)\,(\epsilon(R)-E_2^{(D)})$, 
was derived in Ref.~\cite{rosaBO} solving the energy eigenvalue 
equation~(3) in $D$-dimensions. The solution for $\epsilon(R)$
is given by the transcendental equation
\begin{equation}
2^\frac D2\left(\frac{\sqrt{|\overline{\epsilon}({\overline R})|}}
{{\overline R}}\right)^\frac{D-2}{2} 
K_\frac{D-2}{2}\left({\overline R}
\sqrt{|\overline{\epsilon}({\overline R})|}\right)\\
-  \frac{\pi\csc\left(D\pi/2\right) }{\Gamma(D/2)} \,
\left(1-|\overline{\epsilon}
({\overline R})|^\frac{D-2}{2}\right) = 0\,,
\label{fullbopot}
\end{equation}
where $\overline{\epsilon}(\overline{R}) = \epsilon(\overline{R})|/|E_{2}|$,
being $E_{2}$ the heavy-light bound 
state energy and ${\overline R}$ the dimensionless heavy-heavy  
distance ${\overline R}=R/R_B$, with $R_B = 
\sqrt{\hbar^{2}/{2\mu_{B,AA}|E_{2}|}}$. $K_\alpha(z)$ 
and $\Gamma(z)$ are the modified Bessel function of the second 
kind and the gamma function, respectively.

The full wave function, $\phi(\textbf{R})$, can be separated into a radial
part and an angular part via $\phi(\textbf{R}) = \phi(R)Y_{\textbf{L}}(\hat{R})$, where $Y_{\textbf{L}}(\hat{R})$ are the hyperspherical harmonics~\cite{HAMMER20102212}. By introducing the reduced wave function $\chi(R) = R^{(D-1)/2}\phi(R)$ and considering the presence of ($N-2$) light atoms that generate the effective heavy-heavy potential, 
the heavy-heavy energy eigenvalue equation~\eqref{eqheavy} can be 
written in the radial form
\begin{equation}
\left[-\frac{d^{2}}{dR^{2}}+\frac{(D-3+2L)(D-1+2L)}{4R^{2}}- \frac{m_{A}}{\hbar^{2}}\,V_{AA}(R) \right] \chi(R)  = -\frac{m_{A}}{\hbar^{2}} B_N \,\chi(R)
 \, ,
 \label{scaleheavy}
\end{equation}
where $ B_N =-\big(E_{N} -(N-2)E_2\big)$ and $L$ labels the angular momentum in 
integer and non-integer dimensions. In particular, for $D=3$, the centrifugal 
potential ${(D-3+2L)(D-1+2L)}/{4R^{2}}$ reduces to the standard
centrifugal barrier. Besides that, in $D=2$, $L$ is the familiar eigenvalue 
of the transverse component of the angular momentum operator.
We also observe that the threshold for the bound states in the BO 
approximation is $(N-2)\,E_2$, which is the correct 
one for the $AAB$ system, whereas 
for the $AABB$ system the threshold is the energy of 
the $AAB$ bound state, and so on. Taking these features into consideration, 
we will compute bound states beyond three particles only in the ideal 
case $(N-2)m_B\ll m_A$, for which the BO 
approximation is suitable.

The asymptotic form of 
the heavy-heavy effective potential 
is
given by:
\begin{equation}
V^{(0)}_\text{AA}(R) \equiv\lim_{R\rightarrow 0}V_\text{AA}(R)
=  -\frac{\hbar^{2}}{2\mu_{B,AA}} \,\frac{(N-2)\,g(D)}{R^{2}}=
E_2 \,\frac{(N-2)\,g(D)}{\overline R^{2}}
\, ,
   \label{assympsmall}
\end{equation}
where $g(D)$ comes from the solution of the transcendental 
equation for $\,R\ll R_B$, so that, at distances much smaller 
than the heavy-light bound state size, we have:
\begin{equation}\label{eq:gD}
g(D) = \left[-\frac{ \pi \csc(D \pi/2)}
{ 2^\frac{D}{2} \Gamma({D}/{2}) K_\frac{D-2}{2}\big(\sqrt{g(D)}\big) } \right]^\frac{4}{2-D}.
\end{equation}

The large distance tail of the effective potential can be expanded in the form
\begin{equation}
V_\text{AA}^{(\infty)}(R) \equiv\lim_{R\rightarrow \infty}V_\text{AA}(R) =
 (N-2)\,E_{2} \,\,K_\frac{D-2}{2}\left(\overline R \right)  
\frac{2^{\frac{(D+2)}{2}}\Gamma(D/2)\,\overline R \ ^{\frac{2-D}{2}}}{(2-D)\pi \csc({D\pi/2})}\, ,
\end{equation}
which can be expanded as
\begin{multline}
V_\text{AA}^{(\infty)}(R)=
(N-2)\,E_{2}\, 
 {e^{-\overline R} \over \overline R^{\frac{D-1}{2}}}\,\,\frac{2^{\frac{D+1}{2}}\,\Gamma(D/2)}{(2-D)\sqrt{\pi} \csc({D\pi/2})} 
\\
\times \Bigg[
   1
  + (D-3)\Bigg(\frac{D-1}{8\,\overline R} 
   +
\frac{1}{2}(D-5) (D-1) (D+1)
   \left(\frac{1}{8\,\overline R}\right)^2+\cdots\Bigg)\Bigg]
\, .
\label{assymplarge}
\end{multline}

It is interesting to note that for $D=3$, the tail of the potential 
turns to be the Yukawa form with a healing distance $R_B$ given by the size of the 
heavy-light bound state. For noninteger dimensions, the leading asymptotic term 
keeps the exponential damping but a denominator $\overline R^{(D-1)/{2}}$, 
which makes the potential to deviate from the 
standard Yukawa form for $2<D<3$. 

\begin{figure}[H]
\centering
\includegraphics[width=0.7\textwidth]{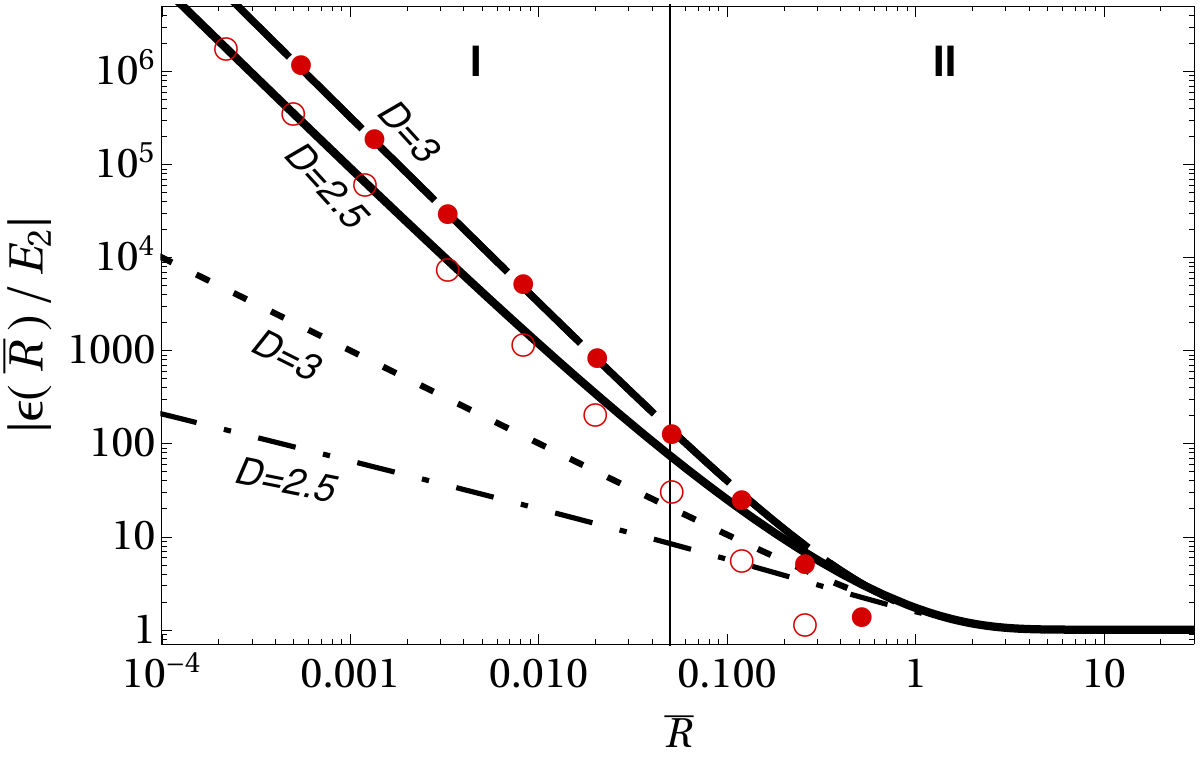}
\caption{Effective potential between the Rubidium atoms for the $^{87}$Rb-$^{87}$Rb-$^7$Li system
as a function of the dimensionless  distance,  $\overline R=R/R_B$ $(R_B=[ \hbar^{2}/{2\mu_{B,AA}|E_{2}|}]^\frac12)$ for the heavy-heavy-light system. The results for the effective potential from the solution of Eq.~\eqref{fullbopot} are presented for $D=3$ (black long-dashed line)
and $D=2.5$ (black solid line).
The  asymptotic form of the potential at large distances $(\overline R\gg 1)$ from Eq.~\eqref{assymplarge} are shown for $D=3$ (black short-dashed line) and  $D=2.5$ (black dot-dashed line). Regions I and II are  separated by the vertical line and  region I corresponds to $\overline R\ll 1$ where the potential is dominated by the asymptotic form  given in Eq.~\eqref{assympsmall}  with 
$g(D=3)=-0.322$ (red solid circles) and with
 $g(D=2.5)=-0.081$ (red empty circles).  }\label{fig2}
\end{figure}

In Fig.\ref{fig2}, we compare 
results for the heavy-heavy effective potential for 
the $^{87}$Rb-$^{87}$Rb-$^7$Li system with finite $E_2$.
We compare the short-distance form of the potential, Eq.~\eqref{assympsmall}, 
with its long-range limit, given by Eq.~\eqref{assymplarge}, 
both written as a function of $\overline R$. The 
short-range part of the potential, 
given by $1/R^2$, leads to
the Thomas-Efimov effect~\cite{PhysRevA.37.3666}. At the unitary limit, region I 
expands to infinitely large distances and the potential takes the form $g(D)/R^2$ 
for all distances between the heavy atoms. We also observe that the tail of the 
effective potential given by Eq.~\eqref{assymplarge} represents it for $\overline 
R\gtrsim 1$ and goes to $E_2$. This exponential damping cuts off 
the excited Efimov state levels that extend much 
beyond the size of the weakly bound heavy-light bound state. It is worth mentioning that a similar discussion was made in 
Ref.~\cite{prafrancisco}, where a different mass ratio was chosen and only 
a part of the long-distance form of the potential 
was explored. 

\subsection{Limit cycles}

\begin{figure}[h!]
\centering
\includegraphics[width=0.495\textwidth]{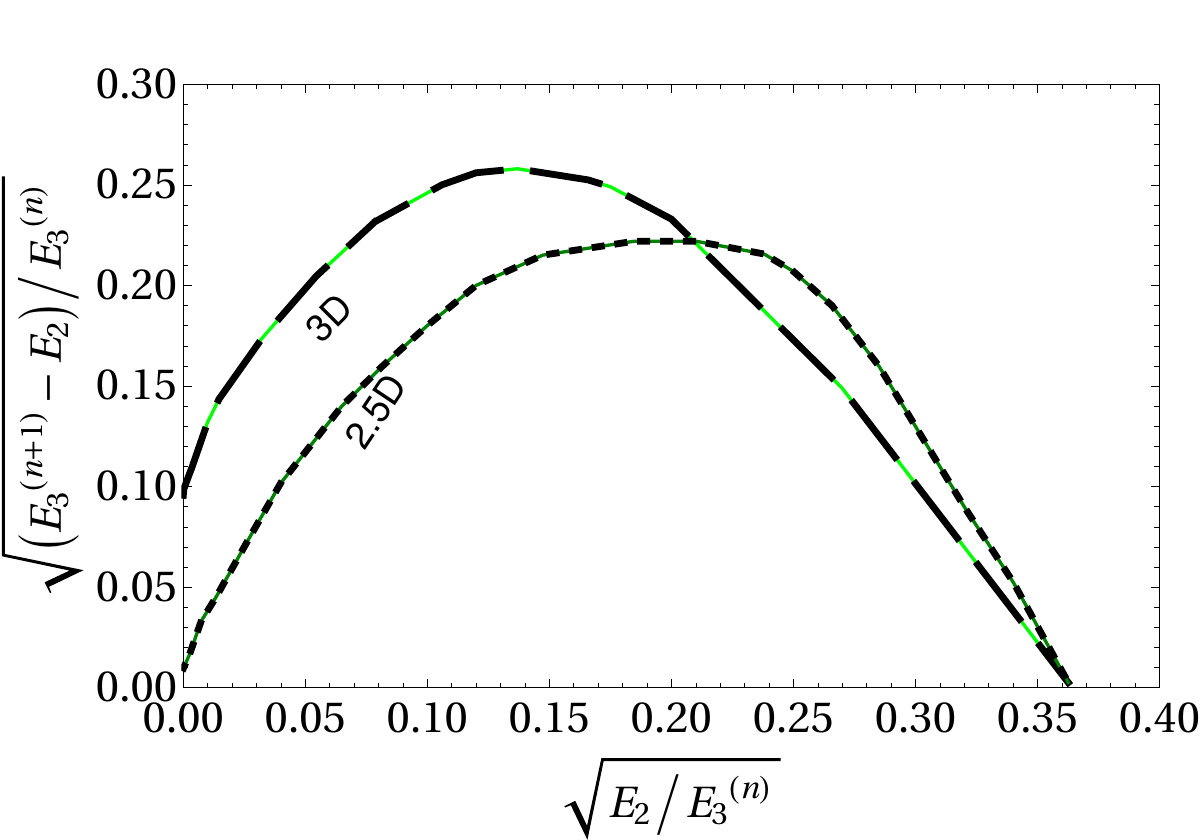}
\includegraphics[width=0.495\textwidth]{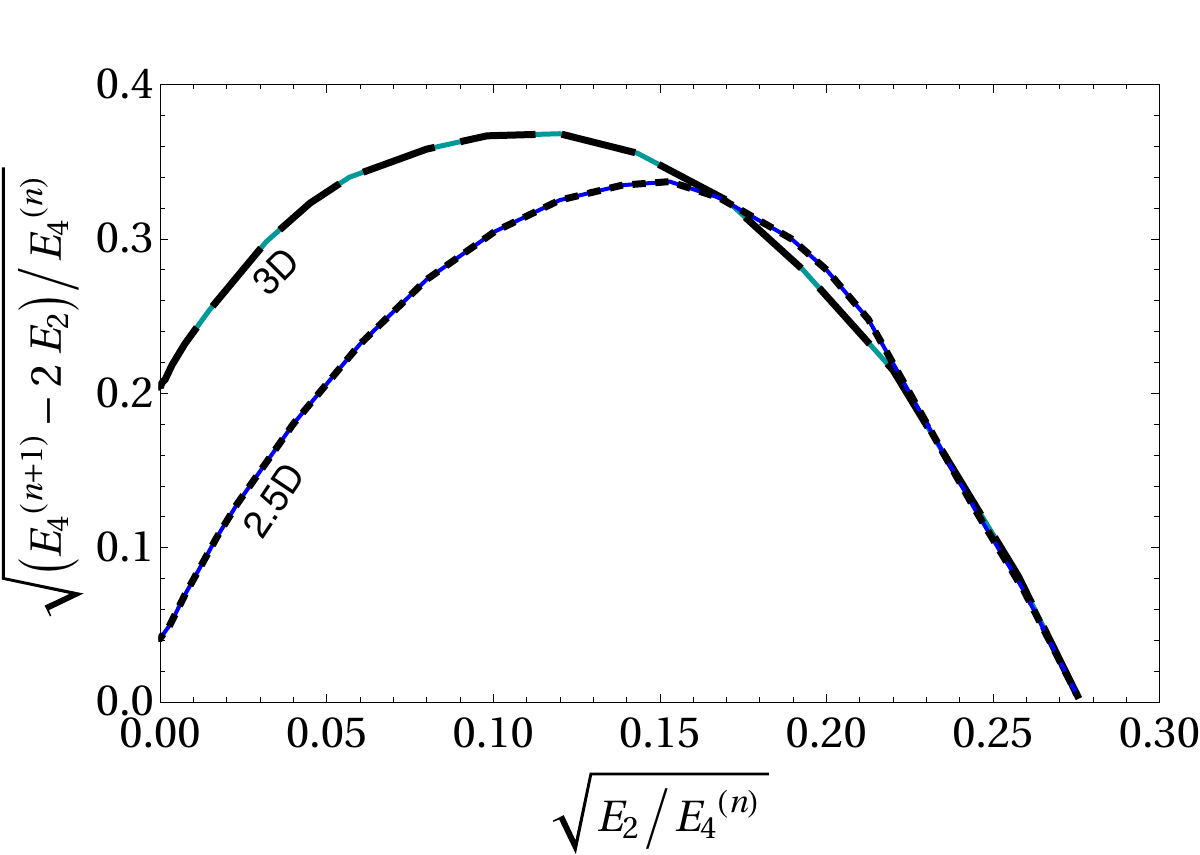}
\caption{Correlations for the three  and four-body bound-state energies of successive  levels, $(n)$ and $(n+1)$ with the ratio $\sqrt{E_2/E_N^{(n)}}$, considering the molecules $^{87}$Rb$_2$-$^7$Li and $^{87}$Rb$_2$-$^7$Li$_2$, represented in the left and right frames, respectively. The systems are embedded in two different dimensions, $D=3$ (long-dashed line) and $2.5$ (short-dashed line). The colored lines represent the successive cycles obtained by starting with the three shallowest levels of the $^{87}$Rb$_2$-$^7$Li and $^{87}$Rb$_2$-$^7$Li$_2$ systems. }
\label{fig:E3E4E2correlation}
\end{figure}

The dependence on the $E_2$ in the three and four-body excited states $(n)$ can be appreciated in Fig.~\ref{fig:E3E4E2correlation}, where it is plotted the correlation between $\sqrt{(E^{(n+1)}_N-(N-2)E_2)/E^{(n)}_N}$ vs. $\sqrt{E_2/E^{(n)}_N}$, obtained with $L=0$ for $N=3$ and 4~(see e.g. \cite{Frederico:2012xh,hadizadeh2011}), computed for the $^7$Li$_{N-2}-^{87}$Rb$_2$ three and four-body systems, respectively. One can observe that, as the ratio between the two-body energy and three or four-body ones increases, the highest excited three or four-body state goes to the continuum threshold. The ratios $\sqrt{E^{(n+1)}_N/E^{(n)}_N}$ at unitarity are such that it is larger for $D=3$ compared to $D=2.5$ for both $N=3$ and 4, which reflects the effect of the dimension in decreasing the strength of the attractive Efimov type $1/R^2$ potential, namely $g(D=3)>g(D=2.5)$. Related to that, the BO approach gives that $g(D)|_{N+1}>g(D)|_{N}$, as it is transparent from the comparison between the ratios of the energies of successive Efimov levels at unitarity  for $N=3$ and 4. The reason for that is traced back to the multiplicative factor $N-2$ entering in the strength of the dominant $1/R^2$ potential for $R<<R_B$, which is given in Eq.~\eqref{assympsmall}. We have computed the correlation plot from the three shallowest states by moving the cut-off radius within the range $0.00001\lesssim R_0/R_B\lesssim 0.1 $, introduced by
$V^{cut}_{AA}(R)= V_{AA}(R)\,\theta (R-R_0)- V_{AA}(R)\,\theta(R_0-R)$ (below $R_0$ the effective potential is repulsive). The limit cycle is achieved fastly and the colored lines represents the next cycle with respect to the long and short-dashed lines.

For a given ratio $\sqrt{E_2/E_N^{(n)}}$, the excited state $n+1$ goes to the continuum threshold, which is represented by $(N-2)E_2$ in the BO approximation. This result is correct for the three-particle system in our example of only heavy-light interaction. As we have already mentioned, in the four-body problem, if $E_3<2 E_2$, the BO approximation is limited in the sense that it allows four body bound states between the light and light-heavy-heavy scattering threshold and light-heavy---light-heavy scattering continuum. 

The unitarity limit configures the situation of the Landau fall to the centre in the heavy-heavy system. This is similar to the Thomas collapse of the system, which can be seen from the heavy particles energy eigenvalue radial equation~\eqref{scaleheavy}. This equation can be written as
\begin{eqnarray}
 &&\left[-\frac{d^{2}}{dR^{2}}+\frac{\Lambda^2-1/4}{R^{2}}\right] \chi(R) = \frac{m_{A}}{\hbar^{2}} E_{N} \chi(R)
 \, ,
 \label{deviationheavy}
\end{eqnarray}
where  the strength of the potential is given by
\begin{equation}
 \Lambda^2 =  -(N-2)\frac{m_A\ g(D)}{2\mu_{B,AA}} + \frac{(D-2+2L)^{2}}{4}\,.
 \label{BOscale}
\end{equation}
The value of $\Lambda^2$ can be either positive or negative. For negative values, we have the Thomas collapse and the log-periodic oscillations in $\chi(R)$, with the wave function exhibiting a discrete scale invariance. In the situation where $\Lambda$ is positive and $R\to 0$, the eigenvalue equation has a continuous scale invariance  and the wave function exhibits a power-law behaviour instead of the log-periodic one. In order to have $\Lambda^2<0$ and a discrete scale invariance, it is necessary to satisfy the following condition:
\begin{equation}
  (N-2)\frac{m_A}{2\mu_{B,AA}}  g(D)>\frac{(D-2+2L)^{2}}{4}\,. \label{eq:thomascollapse}
\end{equation}
The equality on both sides of the above equation defines the critical dimension where the geometrical scaling regime disappears. The above equation shows that it is possible to set a dimension where discrete scale invariance regime is present for the system with $N+1$ particles, while the $N$-body system has continuum scale invariance (see Ref.~\cite{prafrancisco}) - this behaviour shows an interesting independence of $N+1$ and $N$-body scales. 

\begin{figure}[t!]
\centering
{\includegraphics[width=0.495\textwidth]{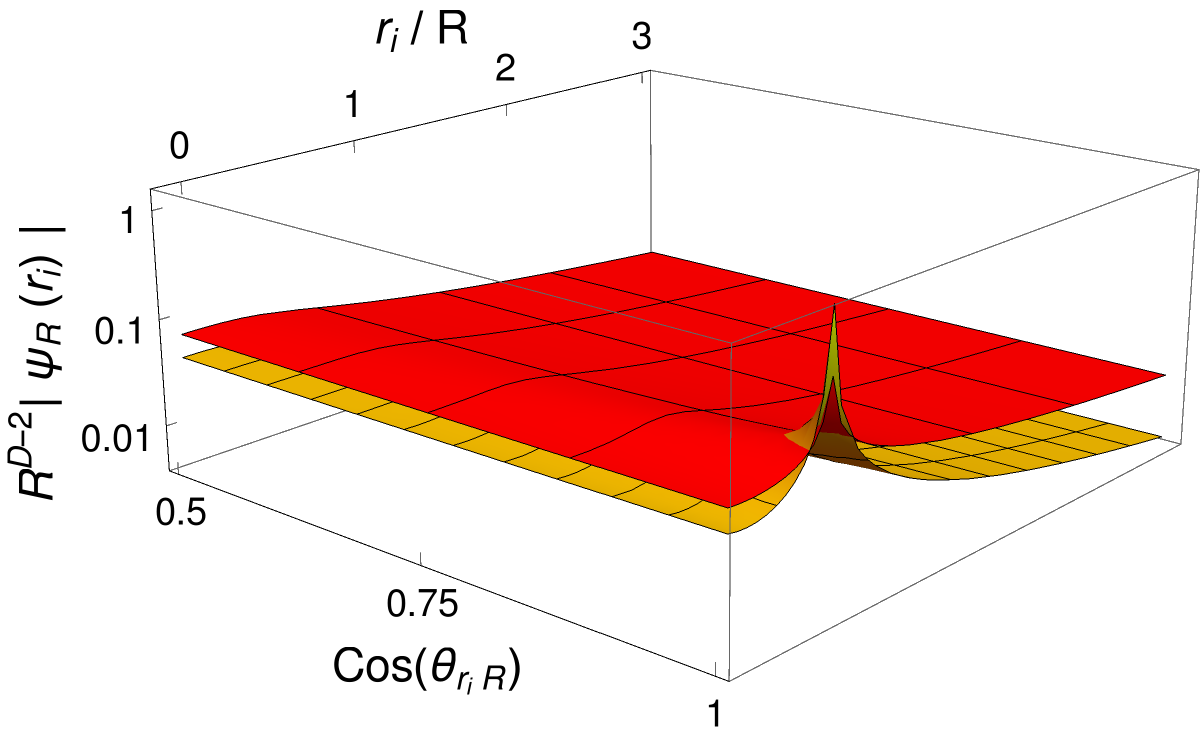}}
{\includegraphics[width=0.495\textwidth]{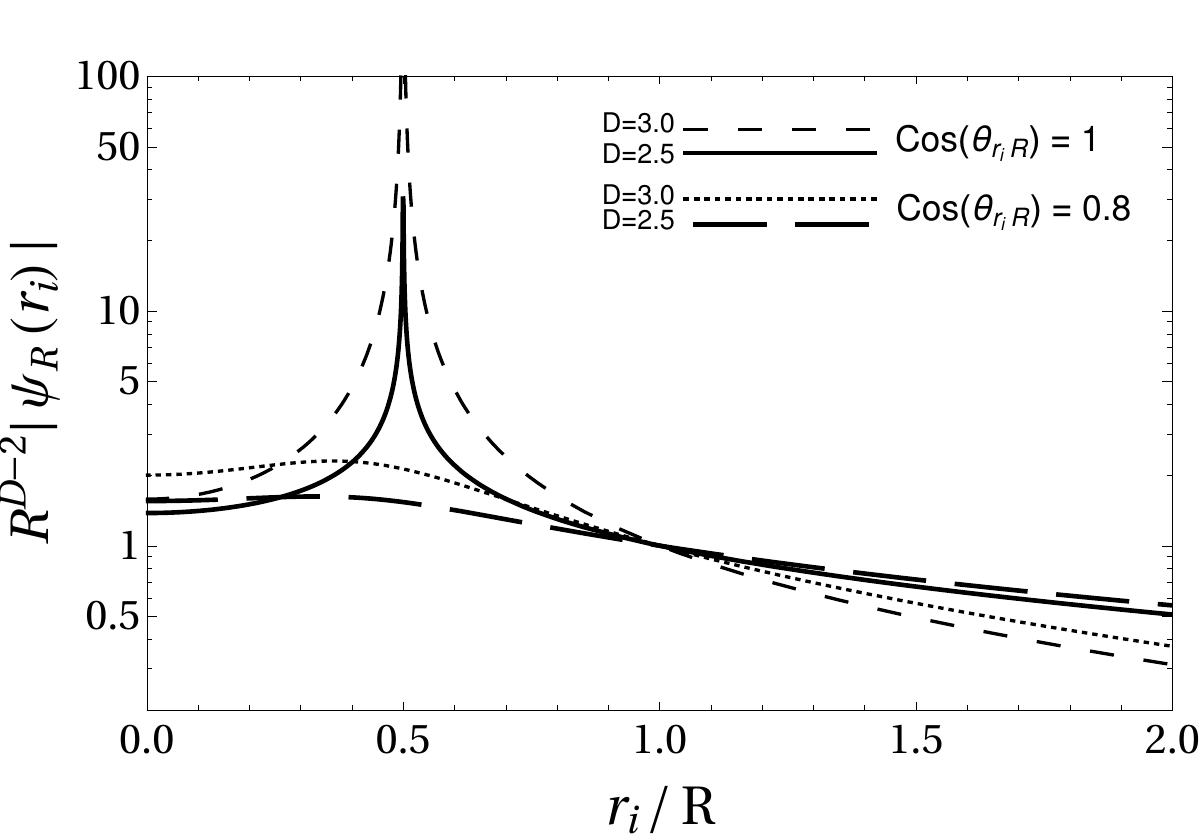}}
\caption{Left panel:  Light-atom wave function in the system $^7$Li$_{N-2}$-$^{87}$Rb$_2$ in $3D$ (yellow surface) and $D=2.5$ (red surface), in the last case, trap geometry $b_{ho}/r_{2D} = 1.414$, the two cases have fixed $ R \sqrt{|E_2|} \equiv 0.0004$ and the used units are $\hbar = m_A=1$.
Right panel: Dependence with $r_i/R$ for
 fixed values of $\{\cos(\theta_{r_i R}),D\}$: $\{1,3\}$ (dashed line), 
 $\{1,2.5\}$ (solid line), 
 $\{0.8,3\}$ (dotted line) and $\{0.8,2.5\}$ (long-dashed line). The wave functions are arbitrarily normalized to $1$ at 
$r_i/R=1$ and $\cos(\theta_{r_iR})=1$.}
\label{fig3}
\end{figure}

\subsection{Light-boson wave function}

The solution of the light-boson energy eigenvalue equation~\eqref{eqlightatom} provides the light-atom wave function $\psi_R(\textbf{r}_i)$ presented in Ref.~\cite{prafrancisco} for finite two-body bound-state energy:
\begin{equation}
\hspace{-.3cm}\psi_{R}(\textbf{r}_{i}) = - \frac{2\mu_{B,AA}}{\hbar^{2}(2\pi)^\frac{D}{2}}
\left(\frac{2\mu_{B,AA}|\epsilon(R)|} {\hbar^{2}} \right)^{\frac{D}{2}-1}
\left[\zeta_{+} ^\frac{2-D}{2} K_\frac{D-2}{2}\left(\zeta_{+} \right)+ \zeta_{-} ^\frac{2-D}{2} K_\frac{D-2}{2}\left(\zeta_{-} \right)\right]. 
\label{fullwavefunciton}
\end{equation}
where
$\zeta_{\pm}=\sqrt{{r_i^2}+\frac{R^2}{4}\pm r_iR\cos(\theta_{r_i R})}\,\sqrt{ 2\mu_{B,AA}|\epsilon(R)|/\hbar^{2}}$.

The results for the light-particle wave function for a system composed by $^7$Li and $^{87}$Rb are presented in Fig.~\ref{fig3}. In the left panel, the peak at $r_i/R=1/2$ and $\cos(\theta_{r_iR}) = \pm1$ corresponds to the light particle at the top of one of the heavy particles. We can also observe that at large distances the light-particle wave function for $D=2.5$ decreases slower than for $D=3$. This behavior follows from the weaker effective potential $\epsilon(R)$ for $D=2.5$ with respect to $D=3$ (cf. Fig.~\ref{fig2}), which leads to a decrease in the light-heavy dimer binding energy when squeezing the system, and therefore resulting in a longer tail for the light-atom wave function. In the left panel of Fig.~\ref{fig3}, for $D$ equal to 2.5 and 3, the behavior of the wave function vs. $r_i/R$, for fixed values of $\cos(\theta_{r_iR})$ equal to 1 and 0.8, are shown. In this example, one can observe that the peak is damped when the angle between the $\vec R$ and $\vec r_i$ increases and the superposition with the heavy particles is avoided. 
The longer exponential tail comes by lowering the dimension, as we have seen already in the left panel.

\subsection{Heavy-heavy wave function}

\begin{figure}[h!]
\centering
{\includegraphics[width=0.45\textwidth]{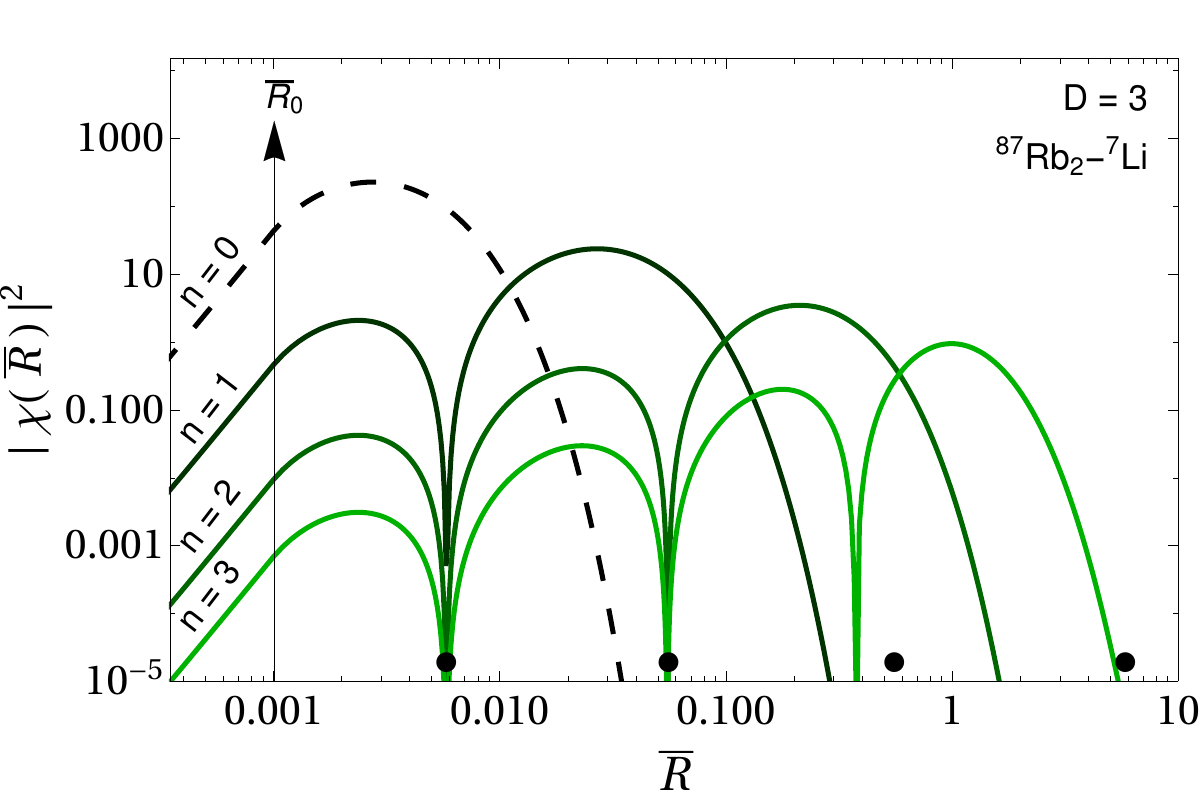}}
{\includegraphics[width=0.45\textwidth]{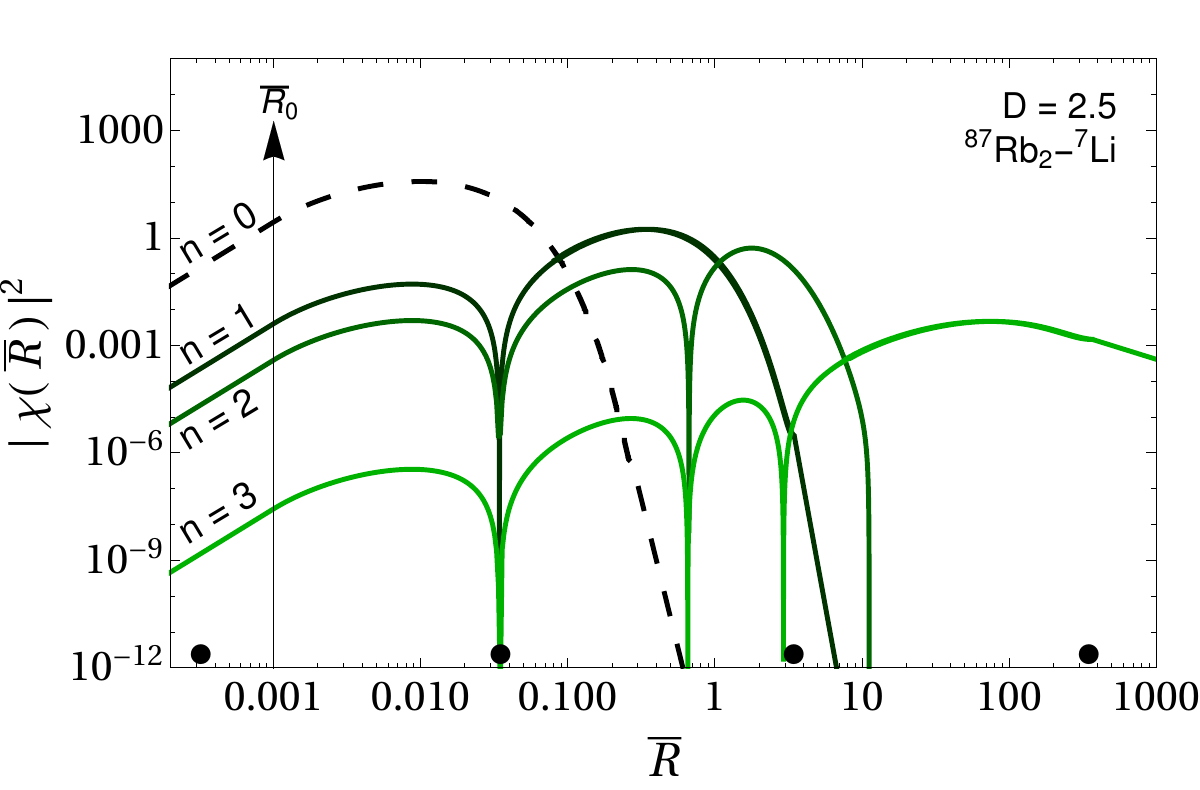}}
{\includegraphics[width=0.45\textwidth]{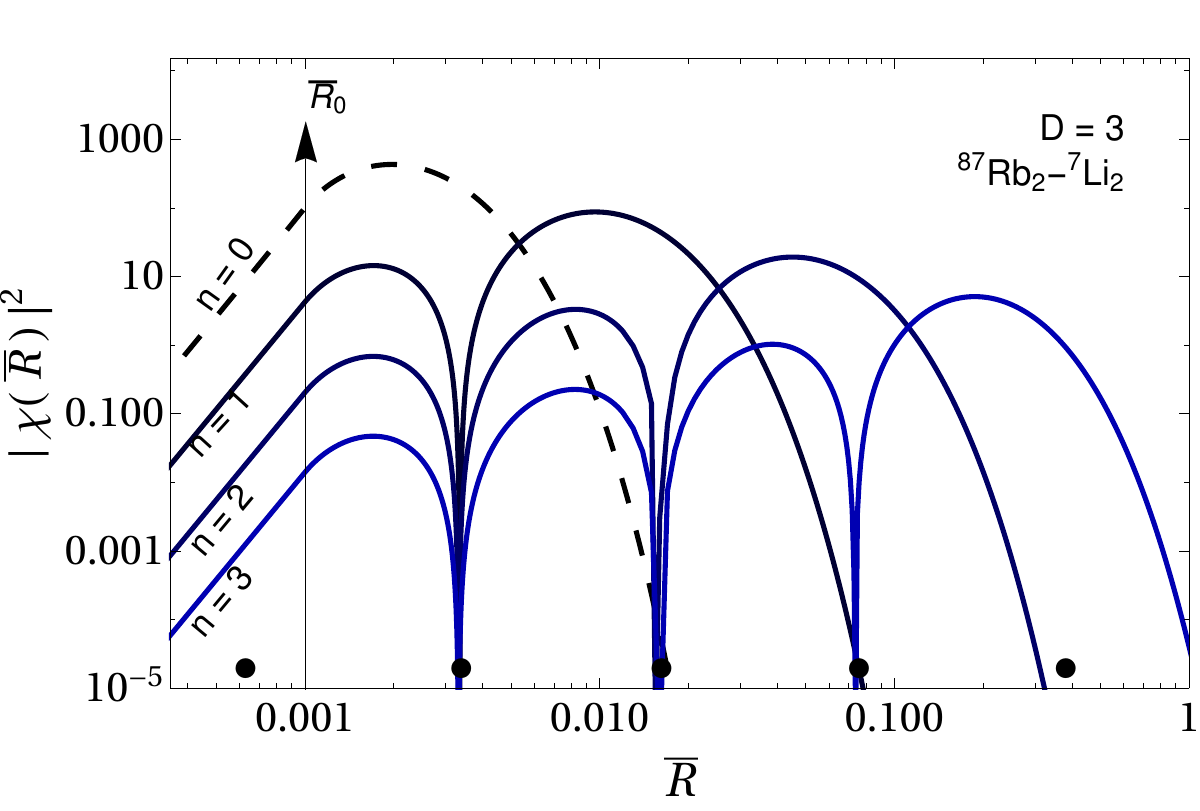}}
{\includegraphics[width=0.45\textwidth]{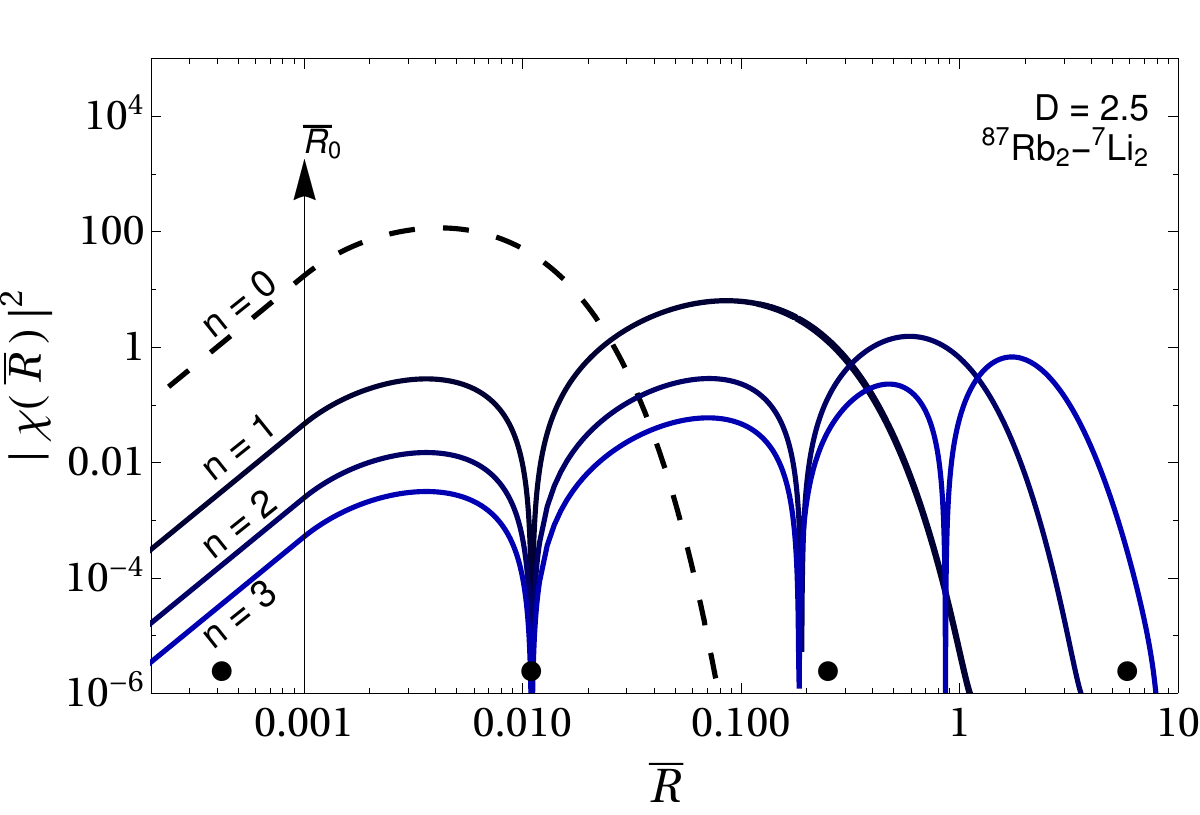}}
\caption{Heavy-heavy bound state radial wave function, $|\chi(\Bar{R})|^{2}$,  for $L=0$ for the  $^{87}$Rb$_2$-$^7$Li three-body molecule (upper panels) and for $^{87}$Rb$_2$-$^7$Li$_2$ four-body molecule (lower panels) embedded in  $D=3$ and $2.5$ presented in the left and right panels, respectively.  The cutoff radius is fixed to $\overline{R}_0 = 0.001$. The dashed lines are the ground state $n=0$,  the solid lines from top to bottom are the $n=1$, 2 and 3 excited states. The dots (black) represent the nodes of the highest excited state when the system is at the unitary limit.}
\label{fig5}
\end{figure}

The heavy-heavy energy eigenstates  for $n = 0, 1, 2$ and $3$ in effective dimensions $D = 3$ and $2.5$ are displayed in Fig.~\ref{fig5} for the  $^7$Li-$^{87}$Rb$_2$ and $^7$Li$_{2}$-$^{87}$Rb$_2$ systems. We have used the heavy-heavy potential with a cut-off radius of $\overline R_0=0.001$ to avoid the Thomas collapse, which damps the wave function for $\overline R<\overline R_0$ as can be observed for both the three and four-body molecules. The exponential damping of the large distance tail comes from the finite binding energy, given by the eigenvalue of the radial Eq.~\eqref{scaleheavy}. This behavior depends on the dimension and number of light bosons, both governing the magnitude of the heavy-heavy adiabatic potential.

For these systems, at the unitary limit, the dots indicate the nodes of the highest excited states, which can be used as a reference on how far from the log-periodicity the states with finite two-body bound energy are. In the left and right panels, we can see that squeezing the system from $D = 3$ to $D = 2.5$, for fixed $\overline R_0$, decreases the logarithmic periodicity. This is due to the lowering of the strength in the $1/R^2$ part of the potential. Besides that, by comparing the heavy-heavy wave function for $N=3$ and $N=4$ molecules, it is possible to observe that by increasing the number of light particles the logarithmic periodicity is enhanced due to the increase of the $1/R^2$ part of the heavy-heavy potential. All these behaviors are in agreement with the previous findings presented in Ref.~\cite{prafrancisco}.

\section{Conclusion}\label{sec:conclusion}

We have studied the discrete scale symmetry regime of the mass imbalanced $^7$Li$_{N-2}$-$^{87}$Rb$_2$ for $N=3$ and 4 systems embedded in different effective dimensions $D$. To accomplish this, we adopted only heavy-light interaction with a contact potential and solved the problem via the Born-Oppenheimer approximation. We discussed the light-particle energy eigenvalue as a function of $D$ and heavy-heavy separation distance, which corresponds to the effective potential in the heavy-particle Schrodinger-like eigenvalue equation. The behavior of the effective potential was studied both for large and small distances between the heavy-particles. As expected, for small distances, the effective potential behaves as the attractive $1/R^2$ with a strength large enough to lead to the Thomas collapse, while for larger distances, its leading asymptotic term displays an exponential damping that turns to be the Yukawa form when 
$D = 3$. The limit cycles, associated with the correlation between the energies of two successive Efimov-like states for the mass imbalanced three- and four-body molecules $^7$Li-$^{87}$Rb$_2$ and $^7$Li$_2-^{87}$Rb$_2$, respectively, were obtained from the BO approximation. The results bring the effect of the finite two-body energy, providing a length scale that cuts the excited state for small enough values, as it is known~\cite{efimov}.

In this work, we have also computed the light-particle wave function and illustrated its behavior, embedding the system in two different dimensions, $D=2.5$ and 3. The highest probability to  find the light particles are at the  top of one of the heavy-particles, so that the wave functions present longer tails and a smoother peak as the effective dimension is diminished. This is associated to the weakening of the effective potential, generated by the light-particles when the non-integer dimension is lowered. 

Finally, for effective dimensions $D = 3$ and 2.5 and energy states $n \,= \,0$, 1, 2 and 3, we have obtained the heavy-heavy wave function in the $^7$Li-$^{87}$Rb$_2$ and $^7$Li$_{2}$-$^{87}$Rb$_2$ molecules. The geometrically separated nodes  of the highest excited states deviate from the log-periodicity found in the unitarity limit due to the damping of the effective potential at the light-heavy finite bound state size. 

For future applications of this model, we intend to study the Borromean situation for negative scattering lengths and analyze how the positions of the resonances~\cite{Bringas:2004zz} and virtual states~\cite{Yamashita:2002zz} are affected by the trap squeezing through variations of the non-integer dimensions.

\section*{Acknowledgments}

This work was partially supported by Funda\c{c}\~{a}o de Amparo \`{a} Pesquisa do Estado de S\~{a}o Paulo (FAPESP) [grant nos. 2017/05660-0 and 2019/07767-1 (T.F.), 2023/08600-9 (R.M.F.), 2023/02261-8 (D.S.R.) and 2018/25225-9 (G.K.)] and Conselho Nacional de Desenvolvimento Cient\'{i}fico e Tecnol\'{o}gico (CNPq) [grant nos. 
306834/2022-7 (T.F.), 302105/2022-0 (M.T.Y.), and  309262/2019-4 (G.K.)].





\bibliography{sn-bibliography}

\end{document}